\documentclass[11pt]{article}

\usepackage{amsmath, amssymb}
\usepackage{graphicx}
\usepackage{geometry}
\usepackage{physics}
\usepackage{hyperref}
\usepackage{natbib}
\usepackage{caption}
\usepackage{subcaption}
\usepackage{booktabs}
\usepackage{float}
\usepackage{tablefootnote}
\usepackage[font=small,labelfont=bf]{caption}
\title{\textbf{Transit Photometry and Ephemeris Refinement of WASP-12\,b Using TESS Data}}

\author{
C.J. Nnaji\\
\small School of Physics, University of the Witwatersrand\\
\small \texttt{(chinedu.jude.nnaji@gmail.com)}
}

\date{}

\begin{document}

\maketitle


\begin{abstract}
\noindent In this study, we conduct a thorough analysis of transit photometry for the ultra-hot Jupiter, WASP-12\,b, using observations acquired by the Transiting Exoplanet Survey Satellite (TESS). The study makes use of light curves and target pixel files, which have been made available through the Mikulski Archive for Space Telescopes (MAST) platform. Following the extraction and normalization of the photometric time-series data, a physical-transit model was applied to the phased light curve to estimate the geometric parameters of the system. Based on the results of the transit model, we estimate the radius ratio, orbital inclination, impact parameter, and duration of the transit. By assuming the stellar parameters reported in literature, we derive the planetary radius and transit depth of the system. We then extract the individual mid-transit times and apply a weighted linear fit to obtain a refined ephemeris. The resulting refined orbital period and reference epoch yield a more accurate prediction of future transit times. A transit-timing-variation analysis was done using an O--C diagram to evaluate the deviation of the timings from a linear ephemeris. There is no evidence for transit timing variations in the dataset used. This study shows how publicly available archival data such as photometry from TESS, and using remote platforms and cloud services (e.g.\ MAST/TIKE), can help analyse and refine the parameters of well-studied exoplanets, which is useful for further investigation of their atmospheres and dynamics, such as in the case of WASP-12\,b.
\vspace{1em}


\end{abstract}

\vspace{1em}

\section{Introduction}

Transit photometry is among the most productive tools used to detect and characterize exoplanets. When a planet crosses the line of sight between the observer and the parent star, it leads to a periodic and observable flux decrease due to the partial obscuration of the stellar disk. The analysis of the resulting transit light curve allows the extraction of many characteristics of the exoplanet and its orbit, such as the planet-to-star radius ratio, the orbital inclination, the impact parameter, and the transit duration, whereas repeated observations allow the refinement of the orbital ephemerides \citep{Winn2010, Seager2010}.

Photometric surveys in space have provided long, precise, and uninterrupted light curves, which have enabled many new discoveries in the field of transit studies. The \textit{Kepler} satellite opened a new era in transit photometry through the discovery of a large variety of planetary systems and the first statistical studies of exoplanet populations \citep{Borucki2010}. Its successor, the Transiting Exoplanet Survey Satellite (TESS), was designed to conduct an all-sky survey of nearby bright stars and provides high-precision transit observations that facilitate follow-up investigations from both ground- and space-based observatories \citep{Ricker2015}. Although TESS observes each field of the sky for a shorter duration than \textit{Kepler}, its photometric precision and ability to monitor target stars over multiple sectors make it capable of improving the orbital parameters of short-period transiting planets.

Among these targets is WASP-12, a bright G-type star hosting an ultra-short-period hot Jupiter, WASP-12\,b. This planet was discovered by the Wide Angle Search for Planets (WASP) survey. It is known for its extremely close orbit, with a period of $\approx 1.09$~days, and its highly inflated radius \citep{Hebb2009}. Due to its proximity to the host star, the planet is subjected to intense stellar irradiation, making it an object of interest for studies of atmospheric physics, tidal interactions, and orbital evolution under extreme conditions. Therefore, WASP-12\,b has been the subject of extensive photometric, spectroscopic, and atmospheric observations \citep{Cowan2012, Sing2013}.

As mentioned above, accurate ephemerides are crucial not only for the physical interpretation of the results but also for observational planning. Small errors in the orbital period can accumulate over time and lead to large errors in predicted transit times, which becomes especially important for short-period exoplanets such as WASP-12\,b. Indeed, the planet completes thousands of orbits even over a few years of observations, and refined ephemerides are useful for future transit observations of the planet, including atmospheric investigations using the \textit{James Webb Space Telescope}. In addition, accurate measurements of transit times allow searches for deviations from strict periodic motion, which may indicate the presence of additional objects or dynamical effects within the system \citep{Agol2005, Holman2005}.

Here we report our analysis of the TESS transit photometry of the exoplanet WASP-12\,b and derive refined parameters of its orbit and ephemeris. We also investigate possible variations in the transit times of the exoplanet.

\section{Observations and Data}

\subsection{TESS Observations}

Photometric observations analyzed in this paper were obtained by the Transiting Exoplanet Survey Satellite (TESS). TESS is a space observatory dedicated to carrying out high-precision time-series photometry of bright nearby stars in order to discover transiting exoplanets \citep{Ricker2015}. TESS observations are divided into sectors, each covering a region of the sky for about 27 days and providing nearly continuous light curves suitable for transit searches, fitting, and timing analyses.

WASP-12 was observed by TESS in multiple sectors during both the primary and extended missions. The ultra-short orbital period of approximately 1.09 days makes the TESS observations rich in transit events and suitable for both detailed transit shape modeling and independent mid-transit time measurements.

\subsection{Data Acquisition and Analysis Environment}

All photometric data were acquired from the Mikulski Archive for Space Telescopes (MAST) through cloud-based access via the TIKE Cloud Science Platform \footnote{\url{https://timeseries.science.stsci.edu/hub/spawn}}. All analysis was performed within this cloud environment, allowing direct access to TESS data products and making the analysis workflow completely reproducible without the need to download data locally.

In this paper, we used Pre-search Data Conditioning Simple Aperture Photometry (PDC\_SAP) light curves produced by the TESS Science Processing Operations Center (SPOC) pipeline \citep{Jenkins2016}. These light curves are corrected for instrumental systematics, spacecraft motion, scattered light contamination, and other time-dependent artifacts affecting the raw photometry. The analysis was performed using TESS Sector~20 observations of WASP-12, obtained with Camera~1 and CCD~3. This sector provides continuous temporal coverage and allows the construction of a high signal-to-noise dataset.

\subsection{Light Curve Preprocessing}

\noindent The downloaded PDC\_SAP light curves were filtered using the quality flags provided with the TESS data products. In particular, cadences affected by known instrumental or spacecraft issues were excluded from the dataset. After quality filtering, the fluxes were normalized to the median out-of-transit value to obtain a dimensionless time series of relative fluxes. All timestamps are reported in Barycentric Julian Date in the TDB time standard (BJD$_{\rm TDB}$).

\begin{figure}[H]
\centering
\captionsetup{width=0.8\textwidth}
\includegraphics[width=0.8\textwidth]{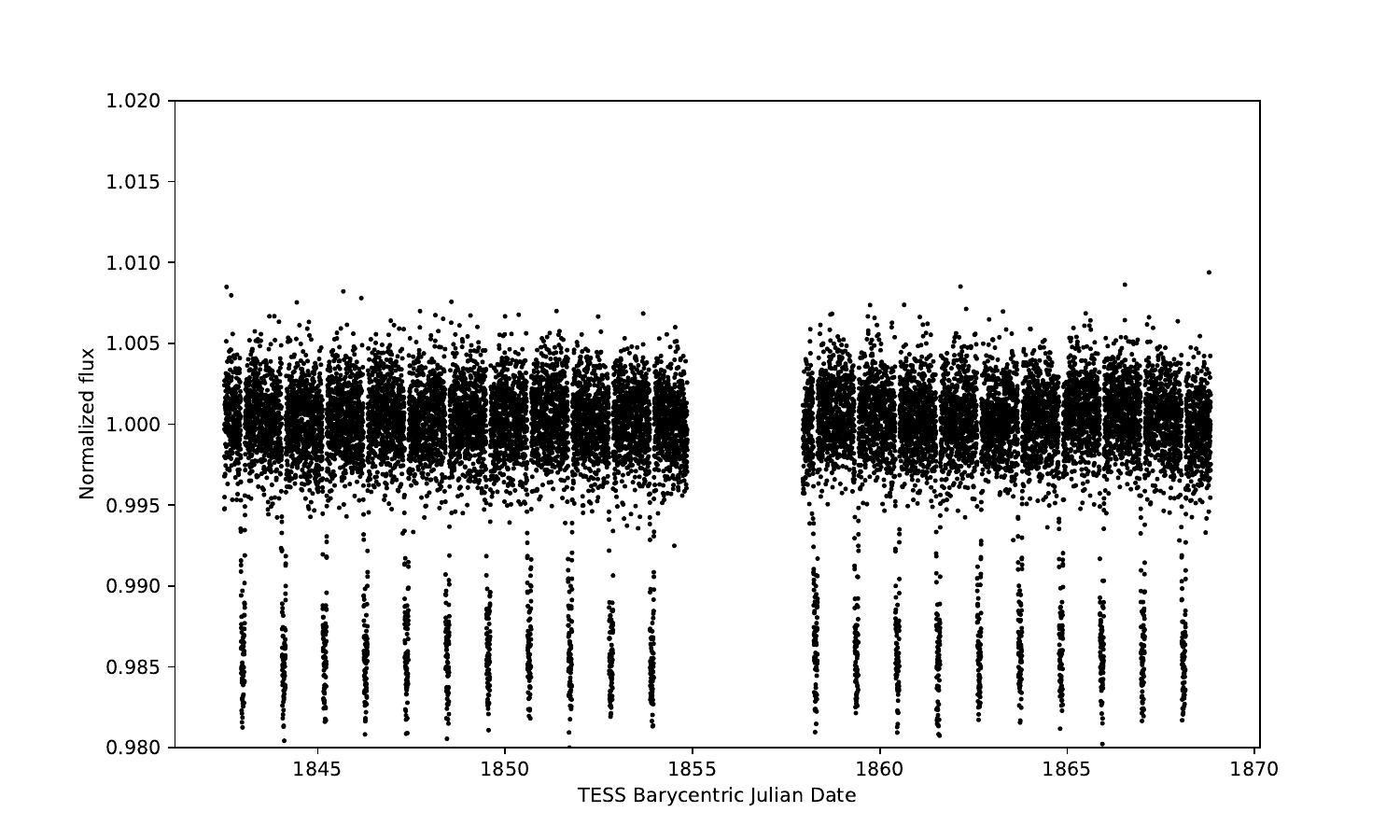}
\caption{Normalized TESS PDC\_SAP light curve of WASP-12. Transit events of WASP-12\,b are clearly visible throughout the observing window.}
\label{fig:full_lc}
\end{figure}

The result of this processing is shown in Fig.~\ref{fig:full_lc}. As can be seen from the figure, transit events of WASP-12\,b are clearly visible in the TESS data.

No further detrending of the light curve was performed.

\subsection{Light Curve Phase Folding and Transit Event Identification}

\begin{figure}[H]
\centering
\captionsetup{width=0.8\textwidth}
\includegraphics[width=0.8\textwidth]{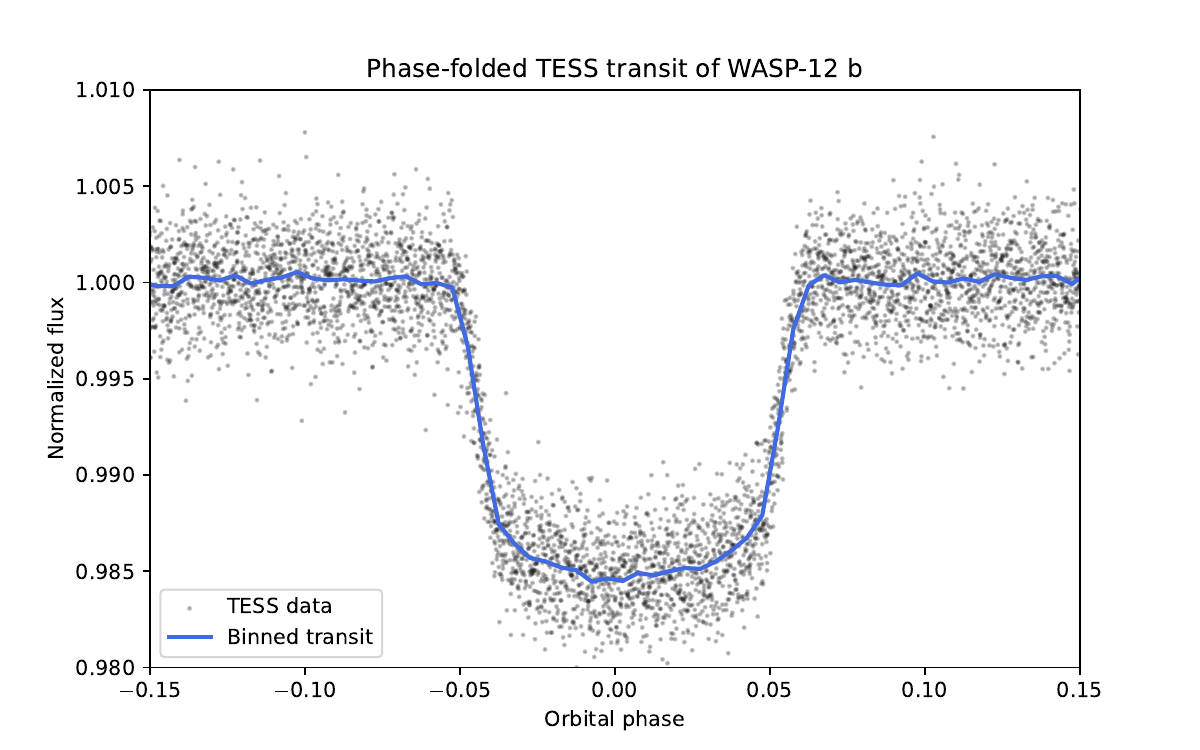}
\caption{Phase-folded and binned TESS light curve of WASP-12\,b using a literature ephemeris. Bin sizes were chosen to enhance the visibility of the transit signal prior to physical modeling.}
\label{fig:phase_folded_binned}
\end{figure}

In order to obtain an initial qualitative assessment of the transit signal, the light curve was phase folded using the orbital period and reference epoch reported in the literature \citep{Hebb2009}. The result is shown in Fig.~\ref{fig:phase_folded_binned} and demonstrates a well-defined transit profile with small scatter.

For visualization purposes, the phase-folded light curve was binned according to orbital phase to reduce photometric noise and emphasize the transit profile. This visualization does not imply any model assumptions; however, it serves as a preliminary step towards the physical transit modeling performed in the subsequent section.

Transit events were identified using the reference ephemeris, and fixed temporal windows centered on the expected mid-transit times were extracted for further modeling and timing analysis.

The resulting dataset was used as the basis for the physical transit modeling and transit timing analysis presented in the following sections.

\section{Methodology}

\label{sec:methodology}

The analysis conducted in this paper uses a step-by-step approach to obtain robust estimates of the transit geometry parameters and to refine the ephemeris of the exoplanet WASP-12\,b using TESS data. The methodology includes physical transit modeling of the phase-folded light curve, as well as an independent timing analysis of the individual transits. All steps of the analysis were performed using Python and the MAST TIKE Cloud Science Platform.

\subsection{Physical Transit Model}

The flux decrement during the transit of a limb-darkened star by a planet was modeled using the \texttt{batman} code \citep{Kreidberg2015}, implementing the analytic transit model by \citet{Mandel2002}.

The model parameters included the planet-to-star radius ratio ($R_p/R_\star$), scaled semi-major axis ($a/R_\star$), orbital inclination ($i$), mid-transit time ($T_0$), and orbital period ($P$). A quadratic limb-darkening law was chosen, with the coefficients set to the literature values corresponding to the stellar parameters of WASP-12. Zero orbital eccentricity was assumed, following previous works on this system \citep{Hebb2009}.

\subsection{Phase-Folded Transit Modeling}

\begin{figure}[H]
\centering
\captionsetup{width=0.8\textwidth}
\includegraphics[width=0.8\textwidth]{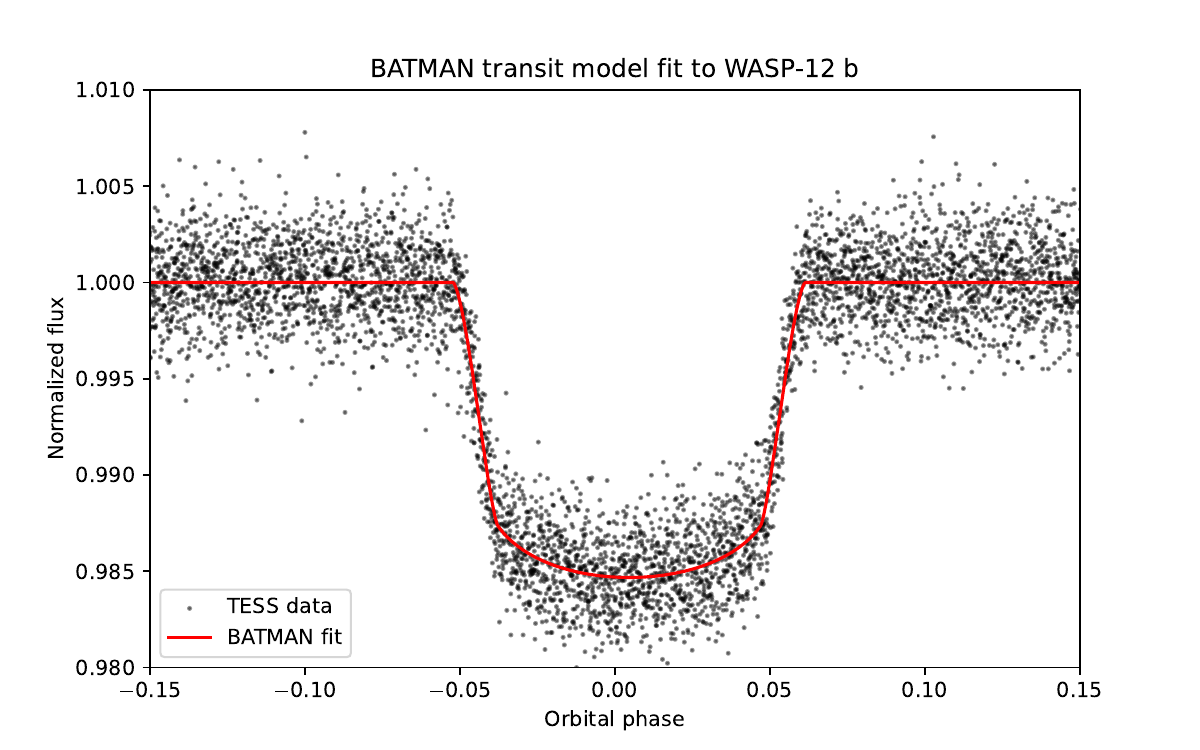}
\caption{Phase-folded TESS light curve of WASP-12\,b with the best-fitting physical transit model.}
\label{fig:batman_fit_no_res}
\end{figure}

The normalized TESS light curve was folded with the initial ephemeris taken from the literature to produce a combined, high signal-to-noise phase-folded transit light curve. This curve was fitted with the physical transit model described above. Minimization of the difference between the observed and model fluxes yielded the best estimates of the transit geometry parameters. The result of the phase-folded fitting is shown in Fig.~\ref{fig:batman_fit_no_res}.

\subsection{Transit Duration and Geometry}

The transit duration ($T_{14}$), which is the time between the first and fourth contacts, was directly computed from the physical transit model. The impact parameter ($b$), which indicates how central the transit is, was calculated using the orbital geometry estimated above. These parameters provide extra constraints on the orbital configuration of the system.

\subsection{Individual Mid-Transit Times}

In order to conduct a timing analysis, mid-transit times were independently measured for each transit event. A fixed window of the TESS light curve, centered on the expected mid-transit time, was selected for each predicted transit. Then, a physical transit model was fitted to this piece of the light curve, with only the mid-transit time and flux normalization varying while keeping other geometric parameters fixed to the estimates obtained at the phase-folded fitting stage. This approach avoids the degeneracy between the transit geometry and its timing \citep{Agol2005, Holman2005}.

Uncertainties of the individual mid-transit times were estimated using the covariance matrix of the fit. Transits with poorly constrained mid-transit times were rejected.

\subsection{Ephemeris Refinement and O--C Analysis}

The refined ephemeris of WASP-12\,b was obtained by fitting the linear relation
\begin{equation}
T_c(E) = T_0 + E\,P
\end{equation}
to the measured mid-transit times, where $E$ is the integer transit epoch. Fitting was done with the weighted least-squares method, where the weight was equal to the inverse square of the measurement uncertainty.

The residuals between the observed mid-transit times and those expected from the ephemeris (O--C) were computed and used to check the deviations from strictly periodic behavior. The statistical significance of the timing residuals was checked with the reduced chi-square test.

\section{Results}
\label{Results}

\subsection{Transit Light Curve and Model Fit}

\begin{figure}[H]
\centering
\captionsetup{width=0.8\textwidth}
\includegraphics[width=0.8\textwidth]{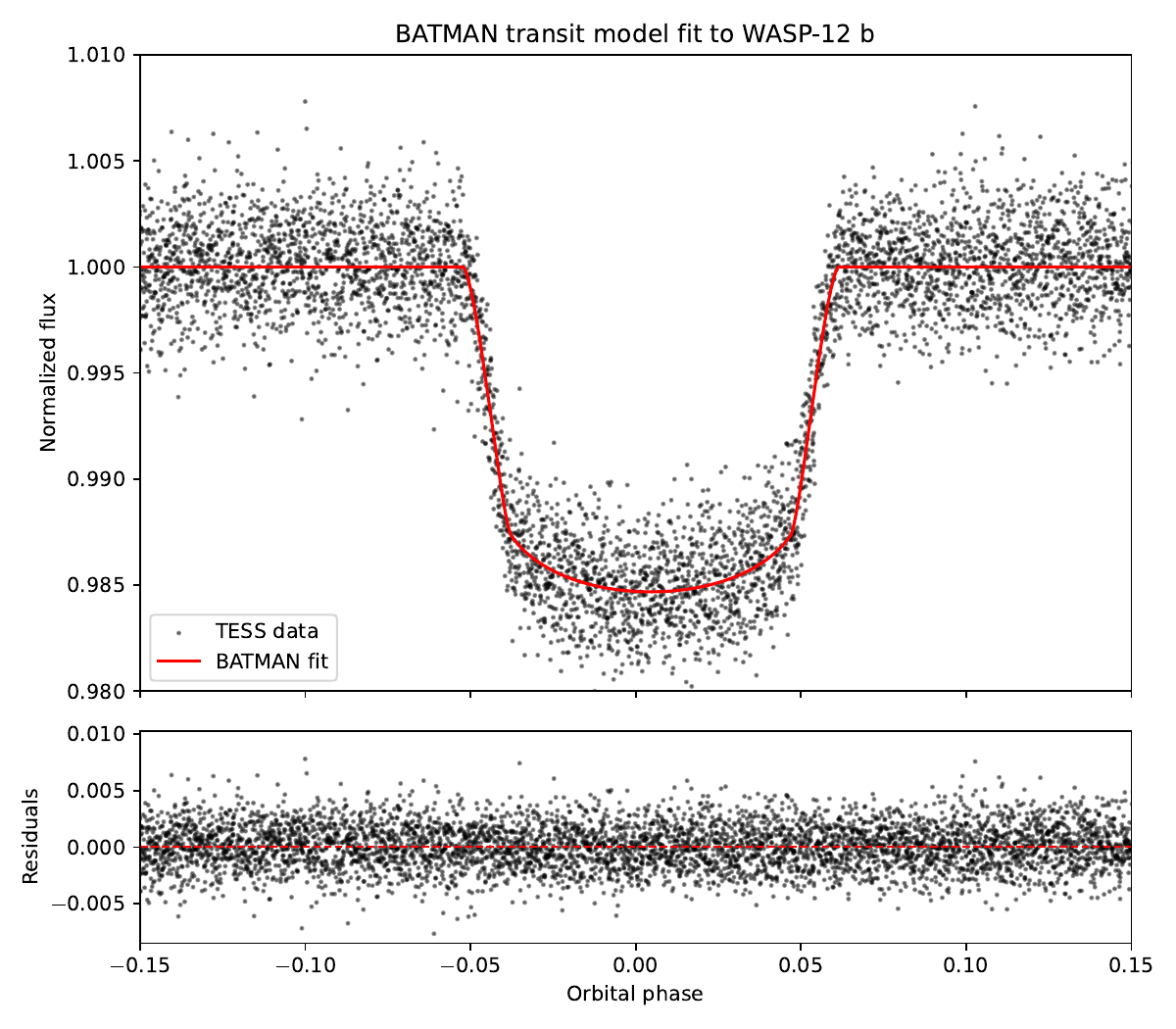}
\caption{Phase-folded TESS light curve of WASP-12\,b with the best-fitting physical transit model (top panel) and residuals (bottom panel).}
\label{fig:batman_fit}
\end{figure}

A physical transit model was fitted to the phase-folded TESS light curve of WASP-12\,b, as detailed in Section~\ref{sec:methodology}. In Fig.~\ref{fig:batman_fit}, the phase-folded TESS light curve of WASP-12\,b is shown together with the best-fitting transit model and the corresponding residuals. The transit light curve exhibits clear transit signatures, and the fitted transit model reproduces the observed ingress, egress, and overall transit shape.

There is no evidence of correlated residuals, indicating that no further detrending or systematic corrections are required for the current analysis.

\subsection{Transit Geometry Parameters}

Key transit geometry parameters of the WASP-12 system were obtained from the best-fitting physical transit model. The planet-to-star radius ratio, scaled semi-major axis, orbital inclination, transit duration, and impact parameter were derived. The transit duration between first and fourth contacts is measured to be

\[
T_{14} = 3.16 \ \mathrm{hours},
\]

which is consistent with previously reported values for this system. The derived impact parameter,

\[
b = 0.74 \pm 0.02,
\]

indicates a moderately non-central transit geometry.

The transit geometry parameters are summarized in Table~\ref{tab:system_params}. The stellar parameters required for conversion to physical quantities were adopted from the literature.

\begin{table}[H]
\centering
\captionsetup{width=0.8\textwidth}
\caption{Stellar and planetary parameters of the WASP-12 system. Stellar parameters are adopted from the literature, while transit and orbital parameters labeled ``This work'' are derived from the TESS analysis presented in this paper.}
\label{tab:system_params}
\renewcommand{\arraystretch}{1.25}
\begin{tabular}{lllll}
\hline\hline
Parameter & Symbol & Value & Unit & Source \\
\hline
\multicolumn{5}{c}{\textbf{Host Star: WASP-12 A}} \\
\hline
Spectral type & -- & G0\,V & -- & \citet{Hebb2009} \\
Stellar mass & $M_\star$ & $1.43^{+0.11}_{-0.09}$ & $M_\odot$ & Exoplanet.eu \\
Stellar radius & $R_\star$ & $1.657^{+0.046}_{-0.044}$ & $R_\odot$ & Exoplanet.eu \\
Effective temperature & $T_{\rm eff}$ & $6360^{+130}_{-140}$ & K & Exoplanet.eu \\
Metallicity & [Fe/H] & $+0.30 \pm 0.10$ & dex & Exoplanet.eu \\
Stellar age & $t_\star$ & $1.7 \pm 0.9$ & Gyr & Exoplanet.eu \\
Distance & $d$ & $432.5 \pm 6.1$ & pc & Exoplanet.eu \\
Visual magnitude & $V$ & 11.69 & mag & \citet{Hebb2009} \\
\hline
\multicolumn{5}{c}{\textbf{Planet: WASP-12 b}} \\
\hline
Orbital period & $P$ & $1.09141782 \pm 0.00002175$ & day & This work \\
Reference epoch & $T_0$ & $2458843.00398 \pm 0.00030$ & BJD$_{\rm TDB}$ & This work \\
Planet-to-star radius ratio & $R_p/R_\star$ & $0.11790 \pm 0.00039$ & -- & This work \\
Transit depth & $(R_p/R_\star)^2$ & $0.01390 \pm 0.00009$ & -- & This work \\
Planetary radius (derived) & $R_p$ & $0.196 \pm 0.006$ & $R_\odot$ & This work \\
Scaled semi-major axis & $a/R_\star$ & $3.022 \pm 0.046$ & -- & This work \\
Orbital inclination & $i$ & $82.54 \pm 0.82$ & deg & This work \\
Transit duration (1--4) & $T_{14}$ & $3.16$ & hr & This work \\
Impact parameter & $b$ & $0.74 \pm 0.02$ & -- & This work \\
Semi-major axis & $a$ & $0.02344 \pm 0.00056$ & AU & Exoplanet.eu \\
Planet mass & $M_p$ & $1.47^{+0.076}_{-0.069}$ & $M_{\rm J}$ & Exoplanet.eu \\
Planet radius (literature) & $R_p$ & $1.90^{+0.057}_{-0.055}$ & $R_{\rm J}$ & Exoplanet.eu \\
Equilibrium temperature & $T_{\rm eq}$ & $2593 \pm 57$ & K & Exoplanet.eu \\
\hline
\end{tabular}

\vspace{0.2cm}
\footnotesize
\textit{Notes.} Stellar parameters are adopted from the Exoplanet Encyclopaedia (\url{https://exoplanet.eu/catalog/wasp_12_ab--459/}) and the discovery paper by \citet{Hebb2009}. Planetary parameters labeled ``This work'' are derived from the TESS transit analysis presented in this study.
\end{table}

\subsection{Revised Ephemeris}

Using the mid-transit times measured from the TESS light curve, a refined linear ephemeris was derived as described in Section~\ref{sec:methodology}. The refined orbital period and reference epoch are

\[
P = 1.09141782 \pm 0.00002175 \ \mathrm{days},
\]

\[
T_0 = 2458843.00397609 \pm 0.00029715 \ \mathrm{BJD}_{\rm TDB}.
\]

These values improve the accuracy of future transit predictions over the TESS observational baseline.

\subsection{Analysis of Transit Timing}

\begin{figure}[H]
\centering
\captionsetup{width=0.8\textwidth}
\includegraphics[width=0.8\textwidth]{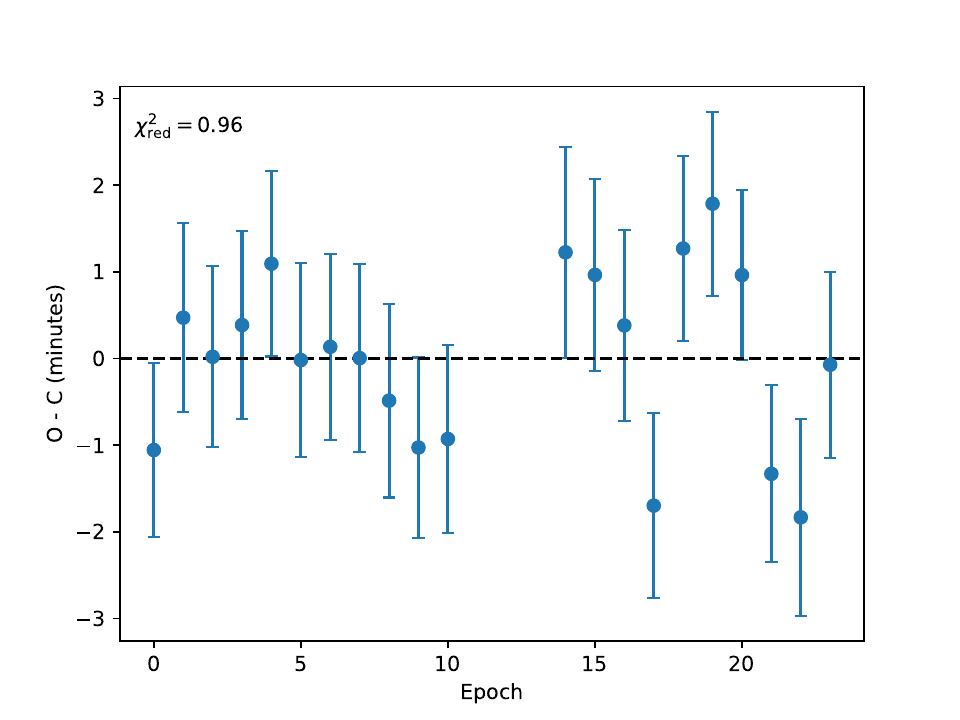}
\caption{Observed minus calculated (O--C) diagram for WASP-12\,b based on TESS transit timing measurements. The dashed line indicates zero timing offset relative to the refined ephemeris.}
\label{fig:oc_diagram}
\end{figure}

The O--C residuals relative to the refined ephemeris are shown in Fig.~\ref{fig:oc_diagram}. The transit timing residuals are symmetrically distributed around zero and have amplitudes of a few minutes. There is no evidence for any periodic trends in the timing residuals.

The reduced chi-square of the ephemeris fit is

\[
\chi^2_{\rm red} = 0.96,
\]

indicating that the scatter in the transit times is consistent with the estimated timing uncertainties. No significant transit timing variations were detected over the TESS observational baseline.

\section{Discussion}

\subsection{Transit Parameters and System Characteristics}

The transit parameters inferred from the TESS light curve data further confirm the unusual nature of the WASP-12 planetary system. Indeed, the inferred planet-to-star radius ratio and transit depth confirm a highly inflated gas giant orbiting very close to its host star. The obtained impact parameter and orbital inclination confirm a transiting system with a moderately grazing transit geometry, naturally accounting for the comparatively long transit duration despite the short orbital period of the planet.

The scaled semi-major axis obtained from the transit analysis confirms the ultra-close configuration of the system. It is known that such close systems are subject to strong tidal and irradiation effects that dominate the evolution of the planet's atmosphere and internal structure \citep{Fortney2007, Cowan2012}. The obtained parameters are consistent with those reported in other ground-based and space-based studies of the system \citep{Hebb2009}.

\subsection{Ephemeris Refinement and Timing Stability}

Based on the transit timings extracted from the TESS photometric data, we obtained a refined linear ephemeris for the WASP-12\,b system. The determined orbital period is consistent with, although more accurately constrained than, previously published estimates.

The O--C diagram shows no deviation from a linear ephemeris. The value of $\chi^2_{\rm red}$ close to unity indicates that the scatter in the O--C diagram is consistent with the estimated uncertainties. Hence, the obtained results provide no evidence for transit timing variations in the TESS data.

The absence of transit timing variations in our study suggests that any additional planetary companion either has a sufficiently small mass or occupies an orbital configuration that does not produce detectable transit timing variations within the sensitivity of the TESS data. This conclusion is consistent with previous analyses of WASP-12\,b, which likewise found no evidence for transit timing variations.

\subsection{Limitations of the Current Study}

The current study is limited by the relatively short baseline and cadence of the TESS data. Transit timing variations with amplitudes below the minute level or with periods longer than the duration of the observations are beyond the sensitivity of this study. Additional limitations may arise from stellar activity and instrumental systematics.

The assumption of a circular orbit is justified for WASP-12\,b based on previous studies; however, the presence of residual eccentricity could introduce systematic effects in the estimated parameters. Further investigations of the system could be performed using radial velocity measurements and multi-band transit observations.

\subsection{Conclusion and Future Prospects}

The refined ephemeris obtained from the TESS photometric data demonstrates the capability of this mission to perform precise analyses of known exoplanetary systems. This work can serve as a foundation for future studies of the atmosphere of WASP-12\,b using facilities such as the \textit{James Webb Space Telescope}.

Continued monitoring of the system over longer baselines, together with the combination of TESS observations, ground-based photometry, and archival datasets from earlier missions, can improve sensitivity to transit timing variations. Such efforts would provide stronger constraints on the presence of additional planets in the system.

\section{Conclusion}

Here we have carried out an analysis of the ultra-hot Jupiter WASP-12\,b based on photometric data collected by the Transiting Exoplanet Survey Satellite (TESS). Using a physical transit model, the phase-folded light curve was fitted to derive the geometric transit parameters, such as the planet-to-star radius ratio, orbital inclination, transit duration, and impact parameter. Utilizing previously determined stellar parameters, we also calculated the planetary radius and transit depth, confirming the highly inflated nature of WASP-12\,b.

An independent analysis of the transit timings was conducted with the aim of measuring the individual mid-transit times and refining a linear ephemeris. The resulting values of the orbital period and reference epoch provide improved predictions of future transit times within the TESS observational baseline. The O--C diagram indicates no statistically significant deviation from a linear ephemeris, suggesting that no transit timing variations are present in the system at the precision level of the data.

The current study demonstrates the continuing importance of TESS photometry for the detailed characterization of known exoplanetary systems beyond the scope of their discovery. Although there is no evidence of dynamical interactions in the WASP-12\,b system, the updated ephemeris and transit parameters can serve as a basis for future atmospheric and dynamical studies.

\section*{Acknowledgment}
This paper includes data collected with the TESS mission, obtained from the MAST data archive at the Space Telescope Science Institute (STScI). Funding for the TESS mission is provided by the NASA Explorer Program. STScI is operated by the Association of Universities for Research in Astronomy, Inc., under NASA contract NAS 5–26555.

\bibliography{references}

\appendix
\section{Additional Figures}
\begin{figure}[ht!]
\centering
\captionsetup{width=0.8\textwidth}
\includegraphics[width=0.6\textwidth]{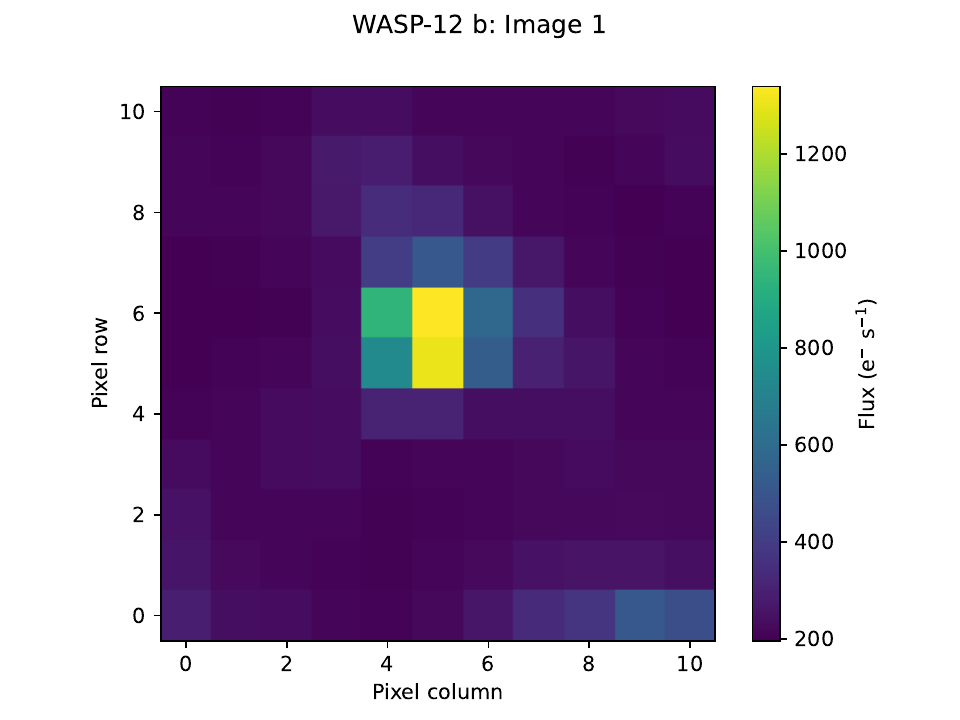}
\caption{
    An example of a target pixel image of WASP-12 obtained from the TESS target pixel files.
    The image shown above corresponds to a single cadence of the calibrated flux data,
    displayed as an $11 \times 11$ pixel cutout centered on the target star.
    The color scale indicates the flux measured in electrons per second.
    The point spread function of the target star is clearly visible, with the flux
    highly concentrated in the central pixels and lower-level background flux present
    in the surrounding pixels.
}
\label{fig:appendix_tpf}
\end{figure}

\begin{figure}[ht!]
    \centering
    \captionsetup{width=0.8\textwidth}
    \includegraphics[width=0.9\textwidth]{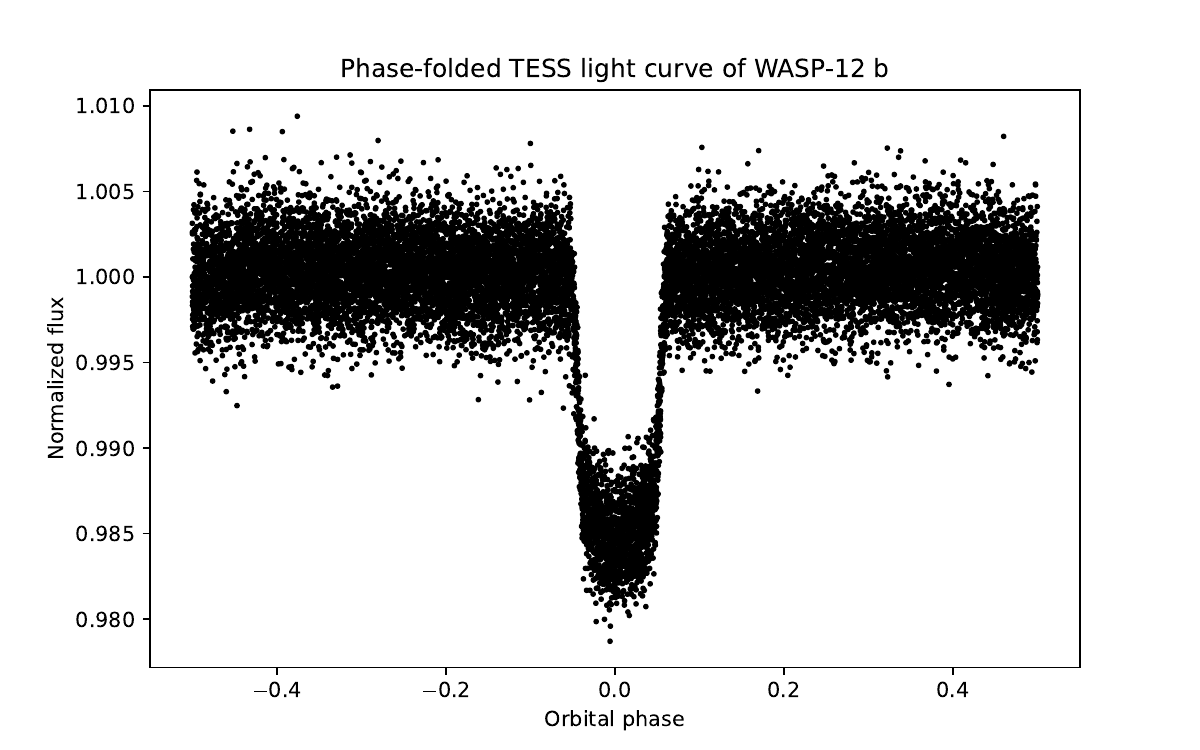}
    \caption{
    Phase-folded TESS light curve of WASP-12\,b based on all available cadence measurements and the reference ephemeris ($P_{\mathrm{ref}} = 1.09142245$\,days, $T_{0,\mathrm{ref}} = 2458843$ BJD$_{\mathrm{TDB}}$).
The light curve was normalized to the median out-of-transit flux level, and no binning or transit modeling
was performed. The figure shows the raw photometric scatter, gaps between observing sectors,
and the complete transit light curve prior to any  binning or model fitting.
The transit depth and overall shape are clearly visible despite the intrinsic photometric variability of TESS.
    }
    \label{fig:appendix_raw_phase_folded}
\end{figure}

\end{document}